\documentclass[preprint,apl,graphicx]{revtex4-1}
\usepackage{newtxtext,newtxmath}
\usepackage{color}
\usepackage[dvipdfmx]{graphicx}

\newcommand{\ket}[1]{\ensuremath{\vert#1\rangle}}
\newcommand{\bra}[1]{\ensuremath{\langle #1\vert}}

\draft 

\begin{document}


\title{
Control of all the transitions between ground state manifolds of nitrogen vacancy centers in diamonds by applying external magnetic driving fields
} 



\author{Tatsuma Yamaguchi}
\email[]{tatsuma317@keio.jp}
\affiliation{School of Fundamental Science and Technology, Keio University, 3-14-1 Hiyoshi, Kohoku-ku, Yokohama 223-8522, Japan}

\author{Yuichiro Matsuzaki}
\affiliation{Device Technology Research Institute,
National institute of Advanced Industrial Science and Technology (AIST),
Central2, 1-1-1 Umezono, Tsukuba, Ibaraki 305-8568, JAPAN
}


\author{Soya Saijo}
\affiliation{School of Fundamental Science and Technology, Keio University, 3-14-1 Hiyoshi, Kohoku-ku, Yokohama 223-8522, Japan}

\author{Hideyuki Watanabe}
\affiliation{Device Technology Research Institute,
National institute of Advanced Industrial Science and Technology (AIST),
Central2, 1-1-1 Umezono, Tsukuba, Ibaraki 305-8568, JAPAN
}

\author{Norikazu Mizuochi}
\affiliation{Institute for Chemical Research, Kyoto University, Gokasho, Uji, Kyoto 611-0011, Japan}

\author{Junko Ishi-Hayase}
\email[]{hayase@appi.keio.ac.jp}
\affiliation{School of Fundamental Science and Technology, Keio University, 3-14-1 Hiyoshi, Kohoku-ku, Yokohama 223-8522, Japan}
\affiliation{Center for Spintronics Research Network, Keio University, 3-14-1 Hiyoshi, Kohoku-ku, Yokohama 223-8522, Japan}


\date{\today}

\begin{abstract}
Nitrogen vacancy (NV) centers in diamonds is
a promising system for quantum information processing and quantum sensing, and the control of the quantum state is essential for practical applications. In this study, we demonstrate a control of all the three transitions among the ground state sublevels of NV centers by applying external magnetic driving fields. To address the states of a specific NV axis among the four axes, we apply a magnetic field orthogonal to the NV axis. 
We control two transitions by microwave pulses and the remaining transition by radio frequency pulses. In particular, we investigate the dependence of Rabi oscillations on the frequency and intensity of the radio frequency pulses. Our results pave the way for a novel control of NV centers for the realization of quantum information processing and quantum sensing.
\end{abstract}


\maketitle 


Nitrogen vacancy (NV) centers in diamonds is
a promising system for the realization of quantum technology that includes quantum information processing, quantum communications, and quantum sensing. 
The established controllability\cite{Dutt2007,Taminiau2014} and long coherence time\cite{Balasubramanian2009,Mizuochi2009,Herbschleb2019} shall be prerequisites for quantum computation. 
The coherent coupling between NV centers and optical photons serves as a method of realizing quantum communication \cite{Bernien2013,Humphreys2018,Yang2016,Tsurumoto2019,Pfaff2014}. 
The spin-spin coupling in diamonds has the potential to realize a quantum simulator as proposed in \cite{Yao2012}. Moreover, owing to the high spatial resolution and high sensitivity at room temperature and atmospheric pressure, one of the most practical applications of NV centers is quantum sensing\cite{Doherty2013,Taylor2008,Rondin2014}. Magnetic fields\cite{Wolf2015a,Taylor2008,Rondin2014,LeSage2013}, 
electric fields\cite{Oort1990,Dolde2011,Dolde2014,Chen2017,Mittiga2018,Michl2019},  and 
temperatures\cite{Acosta2010,Neumann2013,Kucsko,Clevenson2015,Toyli2013,Plakhotnik2014,Wang2018,matsuzaki2016,Hayashi2018} have been measured through NV centers in diamonds.

An NV center in a diamond is a spin-1 system, and there are three states in the ground state manifolds. However, only two of these states are typically controlled, and the remaining state is not addressed using frequency selectivity. Recently, new schemes have been proposed to use all the three states in the ground state manifolds; they have demonstrated high-sensitivity magnetic fields sensing\cite{Fang2013a}, high-sensitivity temperature sensing\cite{Neumann2013}, and AC magnetic field sensing that does not require pulse control\cite{Saijo2018,Yamaguchi2019}. Moreover, the three levels of the NV centers can be used for a quantum memory of a superconducting qubits\cite{Zhu2014,Matsuzaki2015,Kubo2010}.Further developments are expected using the $\Delta$-system\cite{Vepsalainen2019}.

To exploit the full potential of the spin 1 system, it is important to control all the three transitions among the ground state sublevels of the NV centers. Such full control of the three transitions has been investigated by using a combination of magnetic fields\cite{Jelezko2004}, lasers\cite{Yale2013,Golter2014}, mechanical vibrations\cite{Macquarrie2013,MacQuarrie2015,Chen2020}, and electric fields\cite{Klimov2014}. In the existing approach, a complicated fabrication in diamond is required to realize such full control of the three transitions at room temperature because such a scheme requires either mechanical vibrations of the diamond\cite{Macquarrie2013,MacQuarrie2015,Chen2020} or an application of electric fields \cite{Klimov2014}.

In this study, we demonstrate a simple control of the three transitions among the ground state sublevels of the NV centers using driving magnetic fields at room temperature. Importantly, our scheme does not require a special fabrication of the diamond substrate, unlike in the case of the existing scheme\cite{Macquarrie2013,MacQuarrie2015}, which can provide a practical advantage. We implement Rabi oscillations of two transitions using microwave pulses, whereas we drive the remaining transition using radio frequency (RF) pulses. We sweep the pulse duration and detect the state of the spin using optical measurements. In particular, we investigate the dependence of Rabi oscillations on the frequency and intensity of the RF pulses. Further, we compare the experimental results with the theoretical results and observe good agreement between them.

Figure 1(a) details the energy levels of NV centers without an external DC magnetic field. 
The spin 1 system has three states: 
$|0\rangle$, $\ket{ B }  =\frac { 1 } { \sqrt { 2 } } \left( \ket{ 1 } + \ket{ -1 } \right)$,
and $\ket{ D } = \frac { 1 } { \sqrt { 2 } } \left( \ket{ 1 } - \ket{ -1 } \right)$.
NV centers have four possible crystallographic axes, and their levels are degenerate under zero magnetic field. 
To address the states of a specific NV axis, we apply a magnetic field orthogonal to one of the NV axes so that 
the resonant frequencies of the NV centers with the other three axes become far-detuned. 
The orthogonal magnetic field lifts the degeneracy as illustrated in Fig. 1(b), and we obtain $\ket{ B }$, $\ket{ D }$, and $\ket{ 0 }$ as the energy eigenstates of the Hamiltonian. 
Importantly, AC magnetic fields can induce all the transitions between these three states, as we will describe latter. 
This is a stark contrast with the case when $\ket{ 1 }$, $\ket{ -1 }$, and $\ket{ 0 }$ are the energy eigenstates when the applied magnetic field is parallel to the NV axis, which was typically adopted in the previous experiments where we cannot induce the transition between $\ket{1}$ and $\ket{ -1 }$ by applying AC magnetic fields.

 \begin{figure}
 \begin{center}
 \includegraphics[]{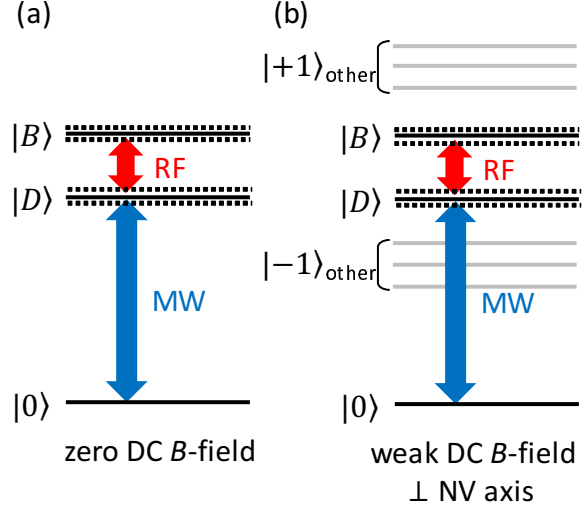}
  \caption{Energy levels of  NV centers (a) Without / (b)With an applied DC magnetic field perpendicular to one of the four possible crystallographic axes. Under the DC magnetic field, we focus on the NV centers with axes perpendicular to the applied DC magnetic field by frequency selectivity. When resonant RF field is applied, Aulter-Towns(AT) splitting occurs and there are four state (dotted lines). The original data is presented in Ref.\cite{Yamaguchi2019}}
 
 \abovecaptionskip=12pt
 \label{NVstate}
 \end{center}
 \end{figure}

We describe the theory about how all transitions between ground state manifolds of NV centers can be controlled by applying AC magnetic fields. Here, let us consider a case to  apply a DC magnetic field orthogonal to the NV axis. The Hamiltonian of the NV center with the magnetic field orthogonal to the NV axis\cite{Yamaguchi2019} is described as

\begin{eqnarray}
H_{\rm NV }=  D{ \hat { S }  }_{ z }^{ 2 }
+ { E }_{ x }\! \left( { \hat { S }  }_{ x }^{ 2 }-{ \hat { S }  }_{ y }^{ 2 } \right)
+ { E }_{ y }\! \left( { \hat { S }  }_{ x }{ \hat { S }  }_{ y } + { \hat { S }  }_{ y }{ \hat { S }  }_{ x } \right)+g\mu ,_bB_x\hat{S}_x,\nonumber \\
\label{Hnv}
\end{eqnarray}
where $ \hat { S } $ is the spin-1 operator of the electronic spin; $ D $ is the zero-field splitting; ${ E }_{ x }$ and ${ E }_{ y }$ are the strains along the $x$ and $y$ directions, respectively; $g\mu _bB_x$ is the Zeeman splitting. The $x$ direction of the NV center is defined as the direction of the applied magnetic field, without loss of generality. Under the condition $D\gg g\mu _bB_x\gg E_y$, we can simplify the Hamiltonian as follows.

\begin{eqnarray}
 H_{\rm{NV}}\simeq D'\hat{S}_z^2 +E_x'\left(\hat{S}_x^2-\hat{S}_y^2\right), \label{effectiveh1}\\
 D'=D+\frac{3}{2}\frac{(g\mu _bB_x)^2}{D+E_x}\\
E_x'=E_x+\frac{1}{2}\frac{(g\mu _bB_x)^2}{D+E_x}
\end{eqnarray}

The eigenstate of the Hamiltonian in Eq. \ref{effectiveh1} is described as 
$|0\rangle$,$\ket{ B }  =\frac { 1 } { \sqrt { 2 } } \left( \ket{ 1 } + \ket{ -1 } \right)$, and
$\ket{ D } = \frac { 1 } { \sqrt { 2 } } \left( \ket{ 1 } - \ket{ -1 } \right)$.
We can control all the transitions between the ground state manifolds of NV centers by applying microwaves and radio frequency fields as follows. The Hamiltonian of the NV center with external magnetic driving fields in our system is given by
\begin{eqnarray}
 H&=&H_{\rm{NV}}+H_{\rm{ex}}\nonumber \\
H_{\rm{ex}} &=& \sum_{j=x,y,z}{{\gamma}_eB}^{(j)}_{\rm MW} { \hat { S }  }_{ j }\cos { \left ({ \omega  }_{ \rm MW }t \right)  } + {{\gamma}_eB}^{(j)}_{\rm RF } { \hat { S }  }_{ j } \cos { \left({\omega}_{\rm RF}t \right) },
\ \ \ \ \ \ 
\label{Hex}
\end{eqnarray}
where ${\gamma}_e$ is the gyromagnetic ratio of the electron spin; ${ B }_{ \rm MW }$ and ${ B }_{ \rm RF }$ are the amplitudes of the microwave and RF fields, respectively; ${ \omega  }_{ \rm MW }$ and ${ \omega  }_{ \rm RF }$ are the driving frequencies of the MW and RF, respectively. 
To control the transition between $\ket{ 0 }$ and $\ket{ B }$ ($\ket{ D }$), we can use microwave driving, as demonstrated in \cite{Dolde2011}. In a rotating frame with $U=-\omega_{\mathrm{MW}}\ket{0}\bra{0}$, the effective Hamiltonian is
\begin{align}
 H\simeq{}&(D-\omega_{\mathrm{MW}}+E_x)\ket{B}\bra{B}\nonumber\\
 &+(D-\omega_{\mathrm{MW}}-E_x)\ket{D}\bra{D}\nonumber\\
&+\tfrac{1}{2}\gamma_eB^{(x)}_{\mathrm{MW}}(\ket{B}\bra{0}+\ket{0}\bra{B})\nonumber\\
&+i\tfrac{1}{2}\gamma_eB^{(y)}_{\mathrm{MW}}(\ket{D}\bra{0}+\ket{0}\bra{D}).
\label{eq:hamiltonianMW}
\end{align}
where we assume that $B_{{\rm{MW}}}\neq 0$ and $B_{{\rm {RF}}}=0$; thus, we can induce Rabi oscillations between $\ket{ 0 }$ and $\ket{ B }$ ($\ket{ D }$) using resonant microwave pulses
with $\omega_{\mathrm{MW}}=D+ E_x$ ($\omega_{\mathrm{MW}}=D-E_x$).

Moreover, we can control the transition between $\ket{ B }$ and
$\ket{ D }$ using radio frequency pulses. In a rotating frame with $U=\frac{1}{2}\omega_{\mathrm{RF}}\ket{B}\bra{B}-\frac{1}{2}\omega_{\mathrm{RF}}\ket{D}\bra{D}$, the effective Hamiltonian is
\begin{align}
 H\simeq{}&(D-\omega_{\mathrm{MW}}+E_x-\frac{1}{2} \omega_{\mathrm{RF}})\ket{B}\bra{B}\nonumber\\
 &+(D-\omega_{\mathrm{MW}}-E_x+\frac{1}{2} \omega_{\mathrm{RF}})\ket{D}\bra{D}\nonumber\\
&+\tfrac{1}{2}\gamma_eB^{(z)}_{\mathrm{RF}}(\ket{B}\bra{D}+\ket{D}\bra{B})\nonumber\\
\label{eq:hamiltonianRF}
\end{align}
where we assume that $B_{{\rm{MW}}}= 0$ and $B_{{\rm {RF}}}\neq 0$; thus, we can induce the coherent oscillation between $|B\rangle $ and $|D\rangle $ with $\omega _{\rm{RF}}=2E_x$.
To our best knowledge, such an experiment to induce the Rabi oscillation between $|B\rangle $ and $|D\rangle $ was not reported before.


\begin{figure}[b]
\begin{center}
\includegraphics[]{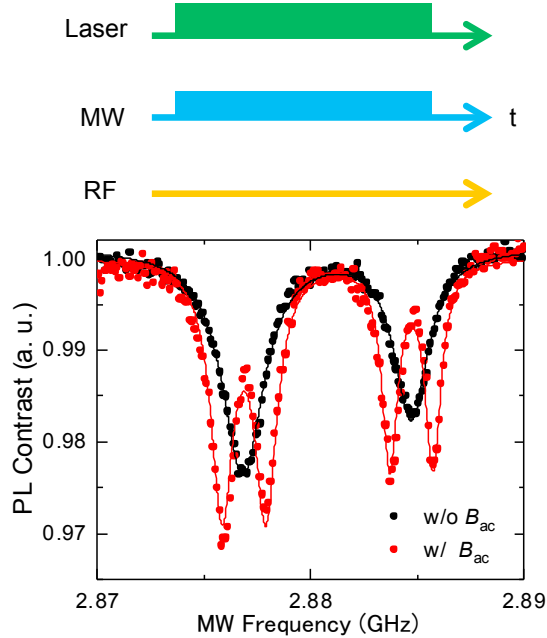}
\caption{Pulse sequence of CW-ODMR without RF fields and CW-ODMR spectrum with and without RF fields under a magnetic field orthogonal to an NV axis.}
\abovecaptionskip=12pt
\label{CW-ODMR}
\end{center}
\end{figure}

We experimentally demonstrate the control of all transitions between ground state manifolds of the NV centers by magnetic driving fields at room temperature.
Our experiment is performed with the same setup and sample as those in Ref. \cite{Yamaguchi2019}. Importantly, although an ensemble of NV centers is used in our experiment, 
we can, in principle, perform the same experiment using a single NV center.
First, we measured the optically detected magnetic resonance (ODMR) spectra under simultaneously applied continuous-wave MW and RF fields, as illustrated in Fig. \ref{CW-ODMR}. In the case without RF fields, two dips are observed in Fig. \ref{CW-ODMR} by sweeping the frequency of MW fields, which correspond to the transitions from $\ket{0}$ to $\ket{B}$ or $\ket{D}$. As a result, the resonance frequency between $\ket{0}$ and $\ket{B}$($\ket{D}$) is estimated to be 2.8768 GHz (2.8847 GHz). From these values, we determined the resonance frequency between $\ket{B}$ and $\ket{D}$, which was estimated to be approximately 7.9 MHz. Furthermore, we measured the CW-ODMR spectrum by applying the RF field with an approximate frequency of 7.9 MHz. As illustrated in Fig. 2, we detected four dips--a clear evidence of the coherent coupling between the RF fields and the transition between $\ket{B}$ and $\ket{D}$. The splitting of each dip corresponds to the AT splitting induced by the RF fields.  Secondly, we demonstrated Rabi oscillations between $\ket{ 0 }$ and $\ket{D}$ ($\ket{B}$) by applying pulse MW fields. As illustrated in Fig. 3, clear Rabi oscillations can be observed by sweeping the pulse duration of MW fields with a resonant frequency between $\ket{ 0 }$ and $\ket{D}$ ($\ket{B}$). The results confirm the $\pi$ rotation between $\ket{0}$ and $\ket{B}$ ($\ket{D}$) with a pulse duration of about 210 ns.

\begin{figure}[h]
\begin{center}
\includegraphics[]{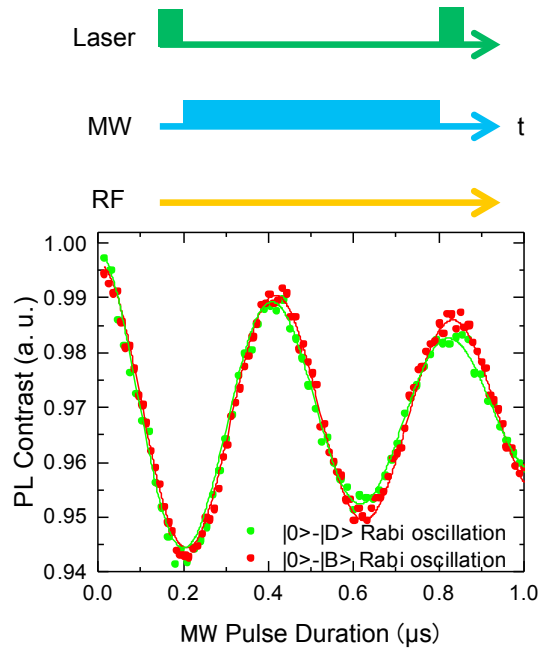}
\caption{Pulse sequences and Rabi oscillations between $\ket{0}$ and $\ket{B}$($\ket{D}$) by applying resonant MW fields. The green plots corresponds to the Rabi oscillations between $\ket{0}$ and $\ket{D}$.The red plots corresponds to the Rabi oscillations between $\ket{0}$ and $\ket{B}$}
\abovecaptionskip=12pt
\label{0Brabi}
\end{center}
\end{figure}

\begin{figure*}[t]
\begin{center}
\includegraphics[]{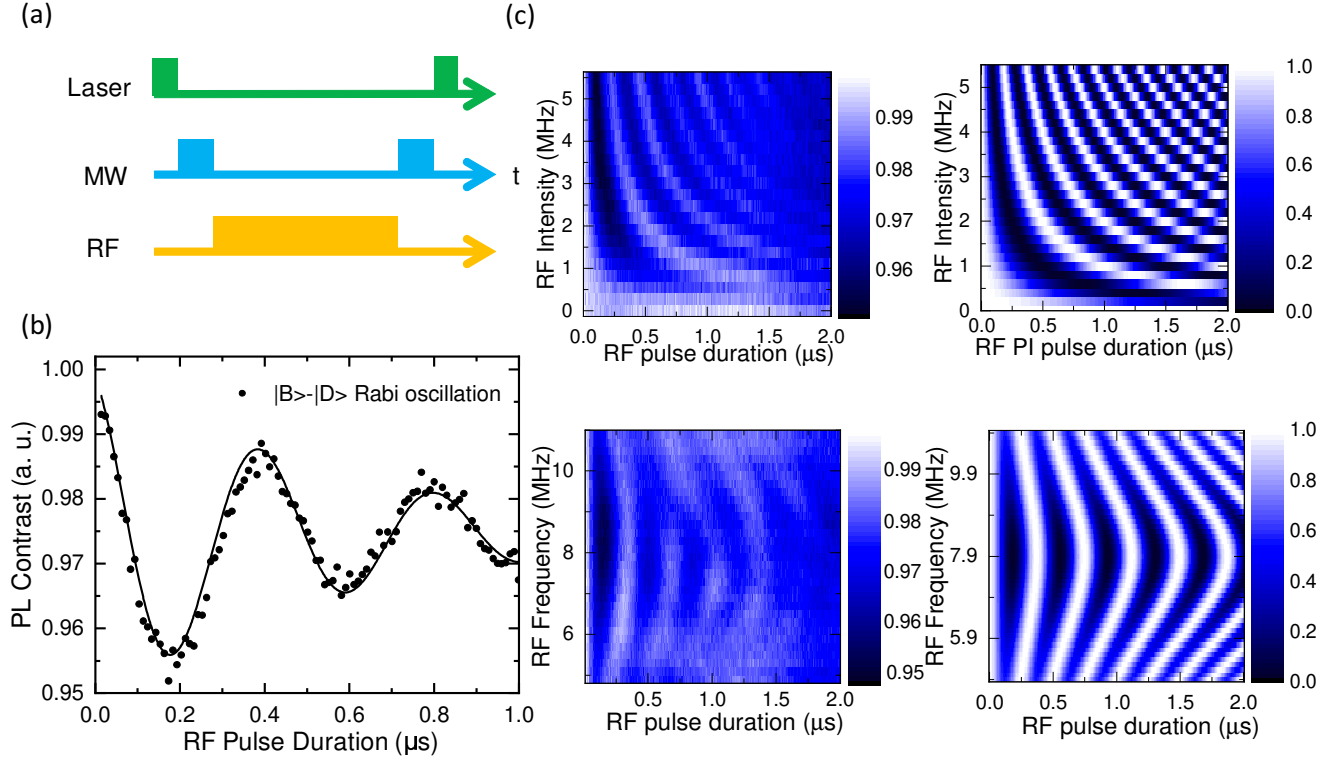}
\caption{Rabi osciiltion between $\ket{B}$ and $\ket{D}$ by resonant RF fields. (a) sequence (b) experimental Rabi oscilation. Rabi oscillations between $\ket{B}$ and $\ket{D}$ by resonant RF fields for various (c) intensities and frequencies of RF fields. The figures on the left and right present the experimental results and theoretical calculations, respectively.}
\abovecaptionskip=12pt
\label{DBrabi}
\end{center}
\end{figure*}

Finally, we demonstrated the Rabi oscillations between $\ket{B}$ and $\ket{D}$. The pulse sequence is presented in Fig. 4(a). It is worth mentioning that the photoluminescence from the state of $\ket{D}$ is the same as that from the state of $\ket{B}$. Thus, to read out the state of $\ket{B}$, we shall convert the population of $\ket{B}$ into $\ket{0}$ by a applying a MW $\pi$ pulse and then read out the state of $\ket{0}$.  The sequence of Rabi oscillations between $\ket{B}$ and $\ket{D}$ is as follows. First, we initialize the NV centers by a green laser, and the state is prepared in $\ket{0}$. Second, we apply the $\pi$ pulse to completely transition from $\ket{0}$ to $\ket{B}$. Third, we implement the RF pulse, where its duration is swept. Finally, we read-out the state of $\ket{B}$ by applying the MW $\pi$ and laser pulses. These experimental results are presented in Fig. 4(b). A clear oscillation can be observed, which confirms the coherent transition between the $\ket{B}$ and $\ket{D}$ states by applying an external magnetic field by the RF pulse. Therefore, we observed all the transitions between the ground state manifolds by applying an external AC magnetic field under a DC magnetic field orthogonal to the NV axis. Furthermore, we measured the $\ket{B}$-$\ket{D}$ Rabi oscillations for different frequencies and intensities of the applied RF pulses, as presented in Fig. 4(c). In the theory, the population difference, w, between $\ket{B}$ and $\ket{D}$ is given by

\begin{align}
 w(t;\Delta)=-1+\frac{2({{\gamma}_eB}^{(z)}_{\rm AC } )^2}{({{\gamma}_eB}^{(z)}_{\rm AC } )^2+\Delta^2}\sin^2{\sqrt{({{\gamma}_eB}^{(z)}_{\rm AC } )^2+\Delta^2}}\frac{t}{2}
\label{eq:rabi}
\end{align}
where $\Delta$ is detuning from resonance frequency. This analytical result as shown in Fig. 4(c) is consistent with the experimental result.

In conclusion, we demonstrated the control of the three transitions among the ground state sublevels of the NV centers by applying external magnetic driving fields and observing the Rabi oscillations between all the transitions. Our technique is significantly simpler than those of previous reports where the mechanical vibration is required to perform quantum manipulation of all the transitions. Owing to the simplicity of our scheme, our results pave the way for the exploitation of the full potential of a spin 1 system.



%
%

%

\begin{acknowledgments}
This work was supported by CREST (JPMJCR1774) and by  MEXT KAKENHI (Grant Nos. 15H05868, 15H05870, 15H03996, 26220602, and 26249108), and QLEAP (No. JPMXS0118067395).
This work was also supported by Leading Initiative for Excellent Young Researchers MEXT Japan and JST presto (Grant No. JPMJPR1919) Japan.
\end{acknowledgments}

\section*{DATA AVAILABLITY}
The data that supports the findings of this study are available within the article.

\bibliography{MyCollection}

\end{document}